\documentclass[letter, traditabstract]{aa}
\usepackage{graphicx}
\usepackage{epsfig}
\usepackage{color}
\usepackage{txfonts}
\usepackage{natbib}
\bibpunct{(}{)}{;}{a}{}{,} 

\newcommand{\Ha}{\mbox{H$\alpha$} }

\def\approxlt{\lower.2em\hbox{$\buildrel < \over \sim$}}
\def\approxgt{\lower.2em\hbox{$\buildrel > \over \sim$}}

\def\gtrsim{\mathrel{\hbox{\rlap{\hbox{\lower4pt\hbox{$\sim$}}}\hbox{$>$}}}}
\newcommand{\kms}{\mbox{{\rm km}\,{\rm s}$^{-1}$}}

\def\lesssim{\mathrel{\hbox{\rlap{\hbox{\lower4pt\hbox{$\sim$}}}\hbox{$<$}}}}

\def\la{\mathrel{\hbox{\rlap{\hbox{\lower4pt\hbox{$\sim$}}}\hbox{$<$}}}}
\def\ga{\mathrel{\hbox{\rlap{\hbox{\lower4pt\hbox{$\sim$}}}\hbox{$>$}}}}

\begin{document}

\authorrunning{Le Tiran et al.}

\title{Can evidence for cosmological accretion be observed in the
H$\alpha$  emission from galaxies at z$\sim$2?\thanks{Data obtained as
part of Programmes ID 074.A-9011, 075.A-0318, 075.A-0466, 076.A-0464,
076.A-0527, 076.B-0259, 077.B-0079, 077.B-0511, 078.A-0055, 078.A-0600,
079.A-0341 and 079.B-0430 at the ESO-VLT.}}

\titlerunning{Can cosmological accretion be observed in H$\alpha$ at z$\sim$2?}

\author{L. Le Tiran\inst{1}, M. D. Lehnert\inst{1}, P. Di Matteo\inst{1},
N. P. H. Nesvadba\inst{2}, \and W. van Driel\inst{1}}

\authorrunning{Le Tiran et al.}

\institute{GEPI, Observatoire de Paris, CNRS, Universit\'e Paris Diderot,
5 place Jules Janssen, 92190 Meudon, France
\and 
Institut d'Astrophysique Spatiale, UMR 8617, CNRS, Universit\'e Paris-Sud, B\^atiment 121, 91405 Orsay Cedex, France}

\date{Received January 25, 2011; Revised April 20, 2011; Accepted April 22, 2011}

\abstract{In previous studies, it has been shown that the large line widths observed in high surface brightness H$\alpha$ emitters at low and high redshifts are likely due to the mechanical energy injected by intense star formation. Here we discuss the possibility that the high surface brightnesses observed are not due to star formation, but due to cosmological gas accretion. We assume that all of the accretion energy is dissipated as shocks from the accreting gas. We show that in order to explain the high surface brightnesses both the mass accretion rate and energy would have to be much higher than expected from simulations or from equating the star formation with the accretion rate. We also investigate scaling relations between the surface brightness expected from accretion and for star formation through mechanical heating and photo-ionization, trying to identify a regime where such accretion
may become evident in galaxies. Unfortunately, the surface brightness
necessary to detect  the gas in optical line emission is about an order
of magnitude lower than what has currently been
achieved with near-infrared observations of distant galaxies.}

\keywords{cosmology: observations --- galaxies: evolution --- galaxies: formation --- 
galaxies: kinematics and dynamics --- infrared: galaxies}

\maketitle

\section{Introduction}\label{sec:intro}

Various processes have been proposed to explain how galaxies obtain the
amount of gas necessary to fuel star formation and their growth. In the
early hierarchical models of galaxy growth \citep[e.g.][]{white78} it
was a combination of minor and major mergers, and gas cooling from a virialized dark matter halo, but it is now thought that much of the cooling may take place outside of the individual halos that subsequently accrete the gas, and that this gas will never be heated up to the virial temperature in the halo -- such accretion mechanisms have been dubbed ``cold flows'' \mbox{\citep[e.g.][]{Dekel09,vdVoort10}}.

All these ideas rely on the basic assumption that baryons are strongly coupled to the dark matter through gravity and thus robustly follow the collapse of the dark matter. Without such coupling and the relative amounts of heating and cooling that this entails, neither cooling
virialized halo gas nor gas from cold flows would deposit as much material
over the same time range as they have been hypothesized to do. However,
there is scant observational evidence for such cold flows or even hot
halos. Recently, \cite{anderson10} concluded that the missing baryons
are not in the halos of nearby galaxies and likely never were part of the
Milky Way. In addition, there is evidence that gas in halos of distant
galaxies is due to starburst driven outflows and not to gas accretion
\citep{steidel10}.

Nevertheless, many models \citep[e.g.][]{Dekel09, Brooks09, Keres09}
suggest that cooling baryons in the halos and/or cold accretion flows
are needed to explain the observations of distant galaxies, especially
their large emission line widths, the apparent prevalence of rotating
disks \citep[as claimed by][]{FS06}, and their high star-formation rates. On the other hand, \cite{L09} (hereafter L09) proposed that the large
line widths observed are not due to gas accretion but to the intense
star formation in the observed high redshift (z$\sim$2) objects. The same conclusion has subsequently been reached by
\cite{Green10}. \cite{Elmegreen10} (hereafter EB10) modeled how accretion
flows might explain the large line widths by dissipating the accretion
energy as turbulence. They conclude that while large line widths are
possible due to accretion, the phase when line widths are dominated by
accretion must be short lived, about one dynamical time ($\sim$100 Myr),
suggesting that intense, or more efficient, star formation may be more likely responsible for the large line widths.

Despite this work, many subsequent studies have continued to suggest that gas accretion is a viable mechanism for explaining some of the properties of high redshift galaxies. In this paper, we examine the two following simple questions: (1.) Can the observed high H$\alpha$ surface brightnesses in distant galaxies be explained by cosmological gas accretion at a rate sufficient to fuel their star formation? (2.) For what ratio of the star formation and accretion rates would we expect the emission line luminosity generated from accretion to dominate? 
The impact of gas accretion can only be observed if the heating rate due
to star formation is below some threshold. This threshold depends on the
efficiency of (1.) the transformation of accretion energy into turbulence,
(2.) gas heating and (3.) the transformation of the mechanical and ionizing
energy of massive stars into turbulent and bulk motions, and heating.

Throughout the paper we adopt a flat H$_0 =$70 $\kms$ Mpc$^{-3}$ concordance cosmology with $\Omega_{\Lambda} = 0.7$ and $\Omega_{M}=0.3$.

\section{\Ha\ luminosity and velocity dispersion \label{sec:2}}
We use ESO archival data from a variety of programs with SINFONI on the ESO-VLT of a sample of more than fifty galaxies in the redshift range 1.3-2.7, which will be described in more detail in Le Tiran et al. (2011, in preparation).  The key to the sample is that all have sufficiently high H$\alpha$ surface brightness to obtain spatially resolved line maps in a few hours integration time. The targets have H$\alpha$ surface brightness levels exceeding a few $\times$ 10$^{-18}$ erg cm$^{-2}$ s$^{-1}$ arcsec$^{-2}$ over more than 1-2 arcsec$^2$ (10-20 kpc) in radius and total H$\alpha$ luminosities above about 10$^{42}$ erg s$^{-1}$. Due to cosmological  dimming, there is a range of a factor of $\sim$5 in the faintest surface brightness line emission levels that can be probed as a function of redshift in our sample. However, it is worth repeating the conclusion from L09 that even our lowest redshift SINFONI sources  are extreme compared to galaxies in the local volume, where only a few of the most intense starbursts have such high H$\alpha$ surface brightnesses, and this on smaller physical scales only.

Many studies which have analyzed the emission-line characteristics of distant galaxies examined velocity dispersions as an integrated property \citep[e.g.][]{Khochfar09, Burkert10, Ceverino10}. These analyses have tended to favor gravity as the source of the observed characteristics, whether driven by gas accretion energy derived from the potential energy of the halo or from gravitational contraction of a gaseous and stellar disk.  L09, on the other hand, focused on the relationship they found between spatially resolved   \Ha\ surface brightness and velocity dispersion in  10 high redshift galaxies observed with SINFONI, a subset of the sample of over 50 galaxies discussed here.  They suggest that neither smooth cosmological gas accretion, velocity dispersions driven by Jeans-unstable clumps, or turbulence generated by energy extracted by a collapsing disk can explain this relationship. They suggested instead that the velocity dispersion due to mechanical energy released by the observed intense star formation  is of the form $\sigma$=($\epsilon\Sigma_{SFR}$)$^{1/2}$, where $\epsilon$ is the efficiency at which the mechanical energy from a star-formation intensity $\Sigma_{SFR}$ is converted into turbulence and bulk flows in the interstellar medium of galaxies. This model has the advantage of having no free parameters since the efficiency can be constrained by observations of nearby galaxies or models \citep[][and references therein]{dib06}.

\begin{figure}
\includegraphics[width=9.5cm]{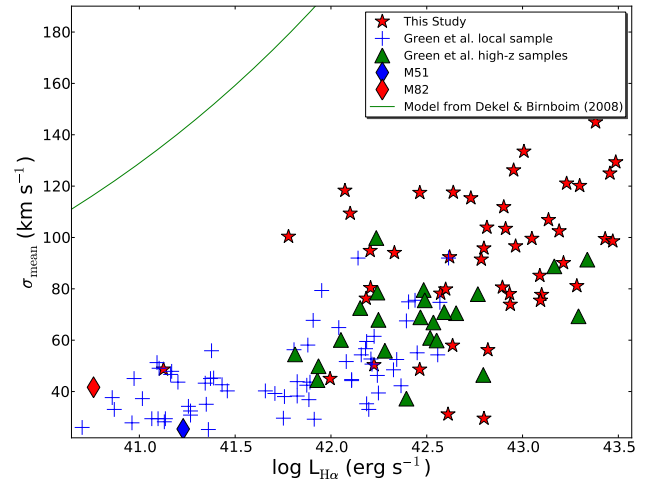}
\caption{Intrinsic mean \Ha\ velocity dispersion (in $\kms$) as function of total \Ha\ luminosity (in erg s$^{-1}$) for different samples of local and high redshift galaxies (see \S ~\ref{sec:2}). The green line represents the H$\alpha$ produced by a gas accretion model \citep[][see \S ~\ref{sec:3}]{dekel08}.}
\label{fig:ha_vs_sigmean}
\end{figure}

Rather than using spatially resolved spectrographic data for a
pixel-by-pixel comparison, \cite{Green10} constructed a single, integrated \Ha\ spectrum per galaxy, through a flux-weighted averaging of their pixels.  Figure~\ref{fig:ha_vs_sigmean} shows the relationship between the total \Ha\ luminosity and the mean velocity dispersion measured using integrated spectra derived with flux weighting per pixel,
for our sample of high redshift galaxies, as well as for the galaxies from \cite{Green10}, i.e., their selection of local SDSS galaxies, M51 \citep{Tully74}, M82 \citep{Lehnert99} and diverse samples of high redshift galaxies taken from \cite{Epinat09}, \cite{Law09} and \cite{Lemoine-Busserolle2010}. The relationship shows a similar trend as the one found between surface brightness and dispersion per pixel in L09.   For our purpose here, this remarkable similarity is important as it supports the notion that any hypothesis that explains the trend of increasing local optical emission line widths with local surface brightness, such as star formation (Lehnert et al. 2009), apparently also explains the similar relationship between  the integrated quantities (something that was not pointed out by either L09 or \citealt{Green10}). We will discuss this in greater depth in Le Tiran et al. (2011, in prep.).

\section{Can cosmological gas accretion shocks produce the
observed \Ha surface brightnesses? \label{sec:3}}

L09 evaluated various mechanisms that can contribute to the large observed \Ha\ line widths, such as clumps, gravitational instabilities, effects from the turbulence and  smooth accretion, and suggested that only the mechanical output from star formation could power such large random motions. This paper further examines the case for cold gas accretion and its observability in the z$\sim$2 Universe \citep[e.g.][]{Keres09, Dekel09}. Gas accretion onto galaxies is undoubtedly a complex process: this accreting gas would further cool through radiation and by generating shocks upon hitting the ISM of the galaxy. The   highly radiative shocks are an effective mechanism through which infalling gas can lose its energy and momentum. But can we see direct evidence for this in our integral field \Ha\ data sets?

The underlying assumption that the infalling gas dissipates energy by colliding with gas already in a galaxy disk and that all the accretion energy, extracted from the halo potential, is converted into supersonic turbulence which then dissipates as fast shocks is 
undoubtedly an over-simplification, as in reality only part of the
energy will go into turbulence and some of it into bulk motions, and shock velocities will span a wide range of velocities which is dependent on the structure (density and temperature) of the ISM. Although some, and perhaps most, of the energy will be dissipated in slower shocks in denser gas, the H$\alpha$ luminosity of the shock will not be very high  in shocks below a few tens of $\kms$ \citep{shull79}.

We can estimate the dissipation rate of the infalling gas using a range of shock models, as we assume that the shock velocities are equal to the observed velocity dispersions, which span the range of $\sim$50-250 $\kms$ in our sample. The shock models are complementary in the sense that some \citep{Allen08} focus on high velocity shocks where the precursor ionizing the gas in front of the shock is important \citep[v$_{\rm shock} \ga$ 80-100 $\kms$,][]{Dopita95}, and others on lower velocity shocks \citep{Raymond79, shull79}.

This allows us to estimate the rate at which mass must pass through the shock in order to provide sufficient H$\alpha$ surface brightness to be observable, as well as the total H$\alpha$ luminosity produced by the accretion energy. 
Using the characteristics of the z$\sim$2 galaxies in our sample, we find a mean velocity dispersion of almost 100 $\kms$ and a total H$\alpha$ luminosity of $\sim$10$^{43.0}$ erg s$^{-1}$. Correcting for the extinction typically observed for this kind of sources would increase their H$\alpha$ luminosities by a factor of 2-5 (L09). Using the shock models of \cite{Allen08}, we find that about 2\% of the shock energy is emitted in H$\alpha$ for a shock velocity of 100 $\kms$, independent of the gas density. The conversion rate  of 2\% is near the maximum efficiency for this process.

An accretion energy extracted from the potential energy of the dark matter halo, of $\sim$10$^{43.1}$ erg s$^{-1}$ has been estimated for a baryonic accretion rate of about 100 M$_{\sun}$ yr$^{-1}$
\citep{dekel08}.   Our typical H$\alpha$ luminosity, if entirely due to star formation, would be consistent with a star-formation rate of $\sim$ 100 M$_{\sun}$ yr$^{-1}$ for a Salpeter IMF with lower and upper mass limits of 0.1 and 100 M$_{\sun}$ \citep[][and a factor of a few lower for a more reasonable Kroupa IMF]{Kennicutt98}. The predicted H$\alpha$ luminosity, if due to accretion, would then be about 10$^{41.4}$ ($\dot{\rm M}_{\rm acc}$/100 M$_{\sun}$ yr$^{-1}$) erg s$^{-1}$, if the whole of the accretion energy is radiated through shocks with velocities of 100 $\kms$, with a conversion rate of 2\%. 
With 10\% of the accretion energy going into supersonic turbulence \citep{Klessen10},
gas accretion rates of many hundreds to thousands of
M$_{\sun}$ yr$^{-1}$ would be required for the total accretion
energy to explain the H$\alpha$ luminosity. Such high values for the gas
accretion rate  are well above what is commonly estimated  from various
models \citep[e.g.][]{Keres09,Genel08}.  We demonstrate this in Figure~\ref{fig:ha_vs_sigmean} where we show the relationship between $\sigma$ and L$_{H\alpha}$ for
a gas accretion model.  For this model, we have used the gas accretion rate as a function of virial velocity and the
energy deposition rate due to accretion \citep{dekel08}, a conversion rate of 2\%, and assumed the virial velocity is equal to the shock velocity.

\begin{figure}
\includegraphics[width=9.5cm]{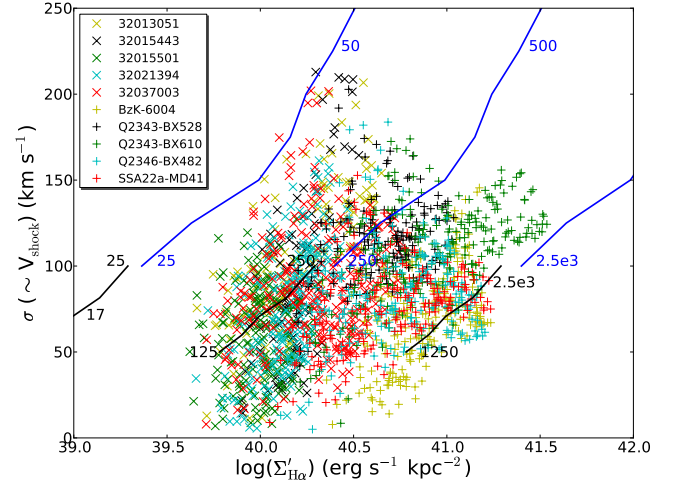}
\caption{Observed pixel-by-pixel H$\alpha$ line widths (assumed equal to
the shock velocities) as function of H$\alpha$ surface brightnesses
(corrected for cosmological surface brightness dimming) in a sub-sample of
5 representative galaxies in each of two redshift bins, z$\sim$1.4  (+
symbols) and z$\sim$2.3 ($\times$). Lines correspond to H$\alpha$ surface
brightnesses  derived using shock models from \cite{Raymond79} (black) and
\cite{Allen08} (blue). The numbers along each line indicate the surface
accretion rates (in M$_{\sun}$ yr$^{-1}$ kpc$^{-2}$) which produce the
H$\alpha$ surface brightness. The 3 sets of lines are for pre-shock
densities of 10, 100 and 1000 cm$^{-3}$, respectively, from left to
right.}
\label{fig:shocks}
\end{figure}

If gas accretion is truly happening at rates of up to several hundreds of
M$_{\sun}$ yr$^{-1}$ at z$\sim$2, it is likely that the integrated
measures of H$\alpha$ luminosity and velocity dispersion are due to a
combination of star formation and gas accretion. We can use the spatially
resolved data to investigate if part of the surface brightness of these
galaxies may be explicable by gas accretion.
For this we will estimate the amount of mass that would need
to be shocked to explain the observed surface brightnesses as a function
of dispersion. 
In Figure~\ref{fig:shocks}, we show the results for 10 representative galaxies in two redshift bins which, because of the strong dependence of surface brightness with redshift, allows us to show the full range of rest-frame H$\alpha$ surface brightnesses within our sample of about 50 galaxies. We find that the surface brightnesses can be explained by shock models, but only if the mass flow rates through the shocks are very high, with   accretion rate densities of over a few hundred M$_{\sun}$ yr$^{-1}$ kpc$^{-2}$ implied for the highest surface brightnesses, whereas only for the lowest
surface brightnesses a more reasonable level of 25-50 M$_{\sun}$ yr$^{-1}$ kpc$^{-2}$ is indicated. In all cases, to explain the surface brightness requires gas of reasonably high density ($\ga$100 cm$^{-3}$).

Taking a rate at which gas is shocked of 300 M$_{\sun}$ yr$^{-1}$
kpc$^{-2}$ would suggest  10$^5$ M$_{\sun}$ yr$^{-1}$ of shock
heated gas, or $\sim$ 10$^{3}$ times the reasonable gas accretion rate \citep{Dekel09}. To support star formation and to form a disk galaxy, the energy from accretion must not be dissipated too rapidly least the gas loses too much angular momentum. Over a dynamical time of about 200-300 Myr (EB10), the amount of gas that needs to be shock heated is of the order of 10$^{13}$ M$_{\sun}$, two orders of magnitude larger than the total ISM mass in a disk galaxy.

\section{Discussion and Conclusions}
The reason why it is likely difficult to note the effects of gas accretion on the ISM of presently observed galaxies is the relatively low efficiency of the conversion of accretion energy into line emission. An interesting question is under what circumstances a galaxy might be observed that is undergoing a (formative) phase of intense gas accretion.

\begin{figure}
\includegraphics[width=9.5cm]{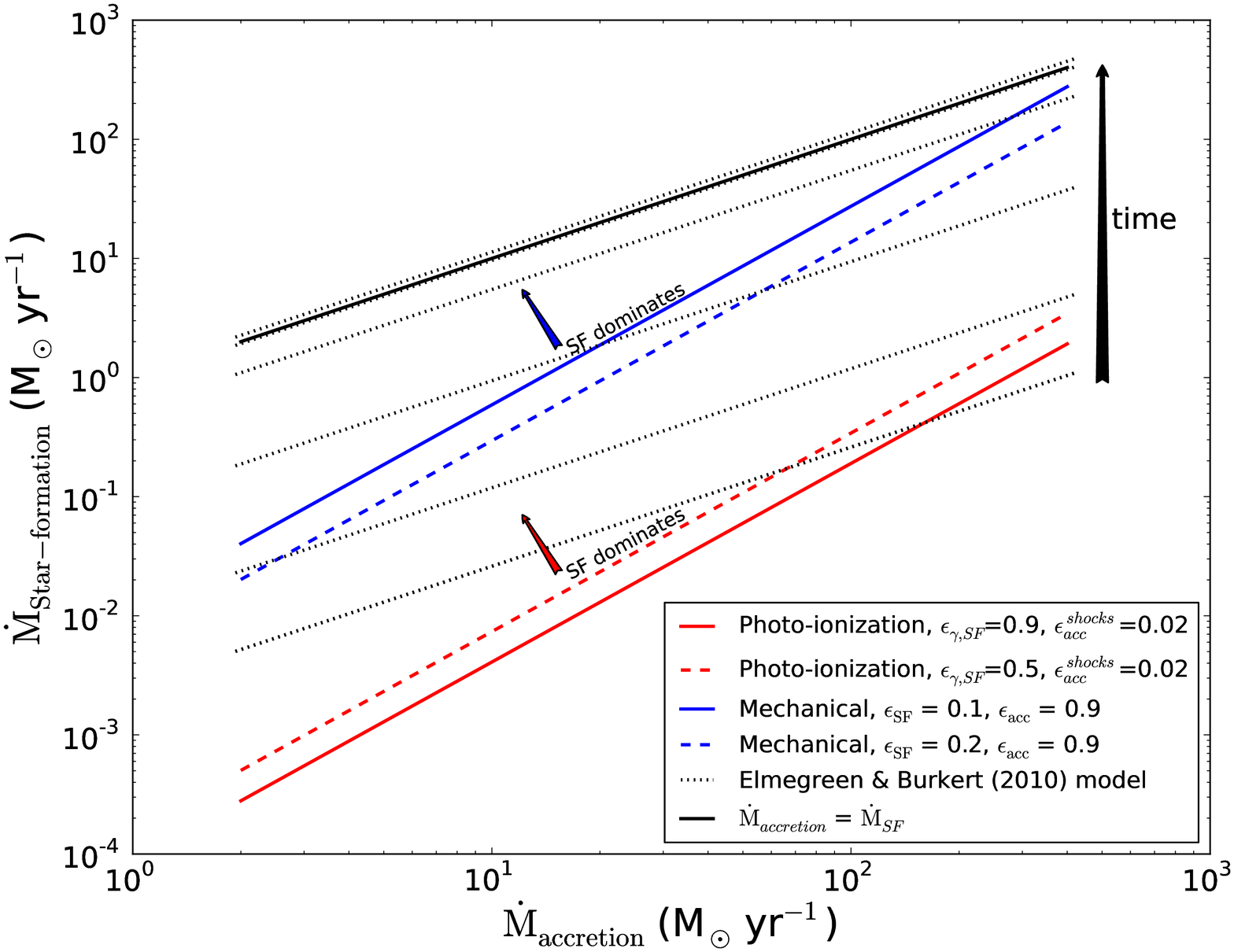}
\caption{Comparison of the star formation rates and cosmological gas mass
accretion rates (in M$_{\sun}$ yr$^{-1}$) at which accretion energy
generates the same H$\alpha$ luminosity as young stars through their
ionizing photons (red solid and dashed lines) and as mechanical energy
released by supernovae and stellar winds (blue lines); the difference
between the two sets of lines is the efficiency at which the ionization or
mechanical energy output couples to the ISM (see text for details). 
Both comparisons assume the same accretion model. Regions above the blue
and red lines are dominated by the influence of the intense
star formation. A comparison with the models by EB10 is
also shown for 0.5, 1, 2, 3, 4 and 5 $\times$ 90 Myr (from bottom to top,
as indicated by the black arrow). The solid black line shows the situation when star
formation and cosmological gas mass accretion rates are equal (which the
model reaches after a couple of dynamical times).}
\label{fig:SFacccomparison} \end{figure}

We have attempted to quantify how the mass accretion rate would compare to the star-formation rate in both controlling the ionization of the gas and in mechanically exciting the gas (see Fig.~\ref{fig:SFacccomparison}). We estimated that the H$\alpha$ luminosity is produced at the rate of 2.6$\times$10$^{41}$ erg s$^{-1}$ M$_{\sun}^{-1}$ \citep[][appropriate for a Kroupa IMF]{SB99}, that there is a 90\% contribution of ionizing photons ($\epsilon_{\gamma, SF}$), which implies an escape fraction of 10\%, that accretion shock energy  is converted into H$\alpha$ luminosity at the rate of 2\% ($\epsilon_{acc}^{shocks}$), and that 90\% ($\epsilon_{\rm acc}$) of the accretion energy is lost due to shocks which emit H$\alpha$ at an efficiency of $\epsilon_{acc}^{shocks}$. To compare this with the mechanical energy of the star formation we used a conversion factor of 7.4$\times$10$^{41}$ erg s$^{-1}$ M$_{\sun}^{-1}$ \citep{SB99} and a conversion efficiency of either 10 or 20\% ($\epsilon_{\rm SF}$).

This allows us to compare the relative efficiency
of heating and ionization from the formation of massive stars with that of the mechanical heating through gas accretion. Can we identify areas in a plot comparing the rates of star formation  and cosmological gas accretion  where observations may be made to discover the duration and impact of gas accretion?  To this end, we have used  the simple model of the relationship between gas accretion and star formation of EB10.  This model, which assumes a phase with constant accretion rate,   allows us to compare, as a function of time, the expected total recombination line luminosity due to infalling gas dissipating its energy through shocks versus that produced by the ionizing radiation and mechanical energy from star formation, whose  rate is estimated by the total gas mass multiplied by the gravitational instability growth rate (see EB10 for details).

In this comparison with the models of EB10 we see (Fig.~\ref{fig:SFacccomparison}) that very quickly, within 45-90 Myr depending on the gas mass accretion rate,
photo-ionization from massive stars dominates over the heating due to
gas accretion. This implies that shock-like line ratios are not expected
to be observed even if the energy of the infalling gas dominates the
dynamics of the ISM. After about 180 Myr, the mechanical energy of massive stars through supernovae and stellar winds begins to dominate, something already concluded by L09. While we should not take the numbers literally, the analysis suggests that the time over which the effects of gas accretion will be evident is relatively short -- less than a dynamical (orbital) time within a galaxy.

So we expect to see the impact of gas accretion on either the ionization
or dynamical state of the gas only early in the gas accretion process
and only when the star-formation rates are relatively low. Within the
context of the EB10 model, the star formation will roughly reach equilibrium with the accretion rate after a couple of dynamical times. If this were the case for our intensely star-forming galaxies at high redshift, we should not see any impact on either the surface brightness or dynamics of the emission line gas, as these are both controlled by star formation, in agreement with what was first proposed by L09 and later supported by the findings of \cite{Green10}.

Can we actually observe this early phase of gas accretion and its impact
on either the ionization or the mechanical energy? The answer is likely
no, at least not through H$\alpha$ observations. Star-forming galaxies
at high redshift have star-formation rates above 10 $M_{\sun}$ yr$^{-1}$
\citep{FS06}. Comparing this with Figure \ref{fig:SFacccomparison} suggests
there is only a small probability that the effect of gas accretion can
be observed. Taking a relatively optimistic star-formation rate of less
than 30 M$_{\sun}$ yr$^{-1}$, which is equivalent to the \Ha\ surface
brightness generated by a gas accretion rate of about 100 M$_{\sun}$ yr$^{-1}$, using our scaling for mechanical energy, suggests that the average surface brightness over 1 arcsec$^2$ (barely enough to resolve a galaxy) would be about 4 $\times$ 10$^{-19}$ erg s$^{-1}$ arcsec$^{-2}$ at z=2.2. This is about an order of magnitude less than what can be detected in a few hours observation with a spectrograph like SINFONI on an 8-10m class telescope like the ESO-VLT, and even beyond the longest integrations made with SINFONI \citep{L10}.   Perhaps more unfortunately, given the difficulty in designing spectrographs with large pixels on the next generation of large telescopes, reaching surface brightness levels beyond the current limits  will remain difficult in the foreseeable future.

Therefore, we conclude that currently, the impact of gas accretion on galaxies at high redshift cannot be observed using the optical emission line gas. This is true especially for the early phases of galaxy formation where the gas accretion rate, despite it being thought of order
100-300 M$_{\sun}$ yr$^{-1}$ for massive galaxies and at high redshifts
\citep{Dekel09, Keres09}, will not be observable both because it does
not liberate enough mechanical energy to be observable with integral
field spectrometers and because it is relatively quickly over-whelmed
by the heating due to star formation, even at relatively modest rates.

\begin{acknowledgements} 

The work of LLT, MDL and PDM is directly supported by a grant from the
Agence Nationale de la Recherche (ANR).

\end{acknowledgements}

\bibstyle{aa}
\bibliographystyle{aa}

\end{document}